\documentclass[a4paper]{spie}  %>>> use this instead for A4 paper
\usepackage[]{graphicx}

\title{RISE: a fast-readout imager for exoplanet transit timing} 

\author{I. A. Steele\supit{a}, 
S. D. Bates\supit{a},
N. Gibson\supit{b},
F. Keenan\supit{b},
J. Meaburn\supit{c},\\
C. J. Mottram\supit{a},
D. Pollacco\supit{b}
and I. Todd\supit{b}
\skiplinehalf
\supit{a}Astrophysics Research Institute, Liverpool John Moores University, CH61 4UA, UK; \\
\supit{b}Astrophysics Research Centre, Queen's University Belfast, BT7 1NN, UK;\\
\supit{c}School of Physics and Astronomy, University of Manchester, M13 9PL, UK
}

\authorinfo{email: ias@astro.livjm.ac.uk}
\begin{document} 
  \maketitle 

%%%%%%%%%%%%%%%%%%%%%%%%%%%%%%%%%%%%%%%%%%%%%%%%%%%%%%%%%%%%% 
\begin{abstract}
By the precise timing of the low amplitude (0.005 - 0.02 magnitude)
transits of exoplanets around their parent star it should be possible
to infer the presence of other planetary bodies in the system
down to Earth-like masses.  We describe the 
design and construction of RISE, a fast-readout frame transfer camera 
for the Liverpool Telescope designed to carry out this experiment.
The results of our commissioning tests are described as well
as the data reduction procedure necessary.  We present light
curves of two objects, showing that the desired timing
and photometric accuracy can be obtained providing that
autoguiding is used to keep the target on the same detector 
pixel for the entire (typically 4 hour) observing run.
\end{abstract}

%>>>> Include a list of keywords after the abstract 

\keywords{robotic telescopes, exoplanets, timing, astronomical instrumentation}

%%%%%%%%%%%%%%%%%%%%%%%%%%%%%%%%%%%%%%%%%%%%%%%%%%%%%%%%%%%%%
\section{INTRODUCTION}
\label{sec:intro}  % \label{} allows reference to this section

The Liverpool Telescope\cite{steele} (LT) is a 2.0 metre 
fully robotic telescope.  The
telescope Acquisition \& Guidance (A\&G) unit can host up to 
five instruments\cite{mottram} with the beam able to be directed
to one of four side ports via a folding mirror, or, with the folding
mirror removed from the beam, to a straight through port.  The rapid
time to switch between instruments ($<30$ seconds) combined with
automated observing scheduling\cite{fraser} means that the telescope
is well suited to time variable (on timescales of minutes
to years) and rapid reaction astronomy.

A key science goal of the Liverpool Telescope is the discovery
of Earth mass exoplanets (i.e. planets outside our own solar system).
Two techniques are pursued.  The first is via participation in the coordinated
followup programmes\cite{robonet} of galactic bulge microlensing
events where small anomalies in the light curves act as signatures of planets
orbiting the lens star.  An example of the application of this technique in which the
LT was involved was the recent discovery of a Jupiter/Saturn solar system analogue\cite{planet}.
The second technique involves the precise timing of transits of known
Jupiter-like exoplanets around their parent star, looking for deviations
from the expected ephemeris due to the influence of other (unseen) planets
in the system perturbing the orbit.  It is this second technique which is
the subject of this paper, which describes the construction of a moderately
wide field, fast-readout CCD imager for the Liverpool Telescope known
as RISE (Rapid Imaging Search for Exoplanets).  In this paper we will give
a brief overview of the science driver for the instrument and the requirements
it places on the design, followed by details of its design and construction.
We will also discuss the data reduction procedure for the instrument, and
give the results of our commissioning tests.

\section{SCIENCE REQUIREMENTS}

The detection of exoplanets is currently of great topical interest  
in astronomy. As technology and techniques progress the main emphasis  
is naturally moving towards the detection of low mass planets. These  
objects are difficult to detect by direct means and it is likely that  
we will have to wait for an ELT to image the first Earth analogue 
system.  However, in the mean time, a number of techniques are being  
used to indirectly detect super-earth massed planets. For an  
eclipsing system, the idea of using timing residuals to infer the  
presence of a third, usually lower mass, body is not a new one,
going back to the discovery of the outer planets of our own solar system.
Recent calculations\cite{holman05,algol05} show that a low mass  
planet moving in a resonance orbit can alter the transit times of a  
hot Jupiter planet by $>5$ sec for periods of 40 days or more. 

Typical transiting exoplanets so far discovered \cite{wasp1,wasp3,wasp4,wasp5} have magnitudes in the range $V\sim8-12$, with amplitudes of 5-20 mmag and
durations of 1-3 hours.  It can be seen that photometry at the 
few milli-magnitude level is therefore required.  Often  
signals such as this are dominated by systematic noise sources.  To
overcome this it is necessary to image as many comparison stars as possible 
with similar brightness to the target star.  Our field of view requirement
($\sim 7.5$ arcminutes) was therefore driven by the maximum field of view
available at a side port of the LT A\&G box.

On a 2-metre telescope the target brightness implies exposure
times of 1-10 seconds to obtain the required SNR for milli-mag photometry.  
This implies a requirement for a readout time of less than 1 second in
order that close to the maximum number of data points (i.e.
total number of photons) is collected over the fixed duration of
the transit.

\section{INSTRUMENT DESIGN AND BUILD}

\subsection{Optical design}

The LT has a plate scale of $\sim 97$ microns/arcsecond, meaning that
to cover a $7.4 \times 7.4$ arcminute field of view a simple focal plane CCD
would be $\sim 46 \times 46$mm.  While such CCD's do exist, they are expensive
and with the large number of pixels implied by the large detector area, 
generally slow to readout with low noise.  It was therefore decided to use 
a smaller, frame transfer CCD combined with a field condensing system 
to achieve 
the combination of field of view and readout speed.

The camera selected was a Andor DW435 model.  This uses an
an E2V CCD47-20 back illuminated, frame transfer CCD which
has 1024 x 1024 pixel light
sensitive region (LSR).  Each pixel is 13.0 x 13.0 microns.  Used in
direct imaging mode such a device would therefore yield a field of view of
2.3 x 2.3 microns with a pixel scale of 0.13 arcseconds per pixel.  A condensing
system with a linear magnification of $\sim 0.3 \times$ was therefore required.

\begin{figure}
\begin{center}
\includegraphics[height=10cm]{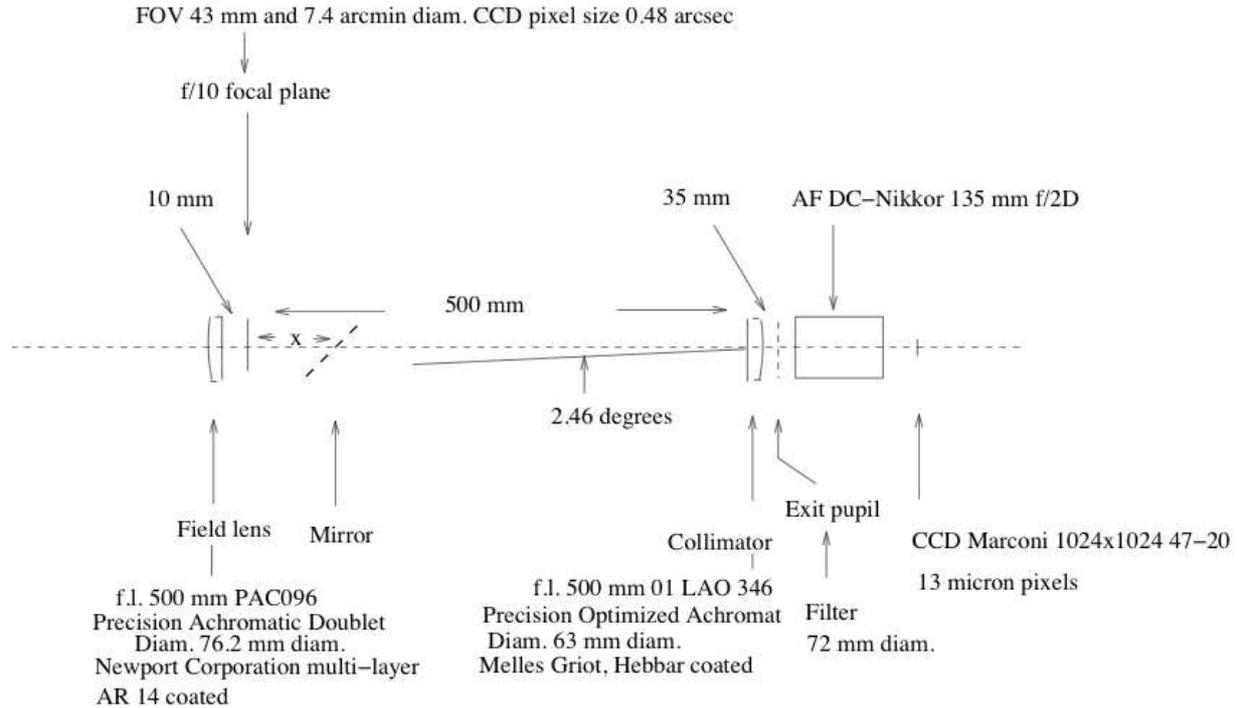}
\end{center}
\caption[]{\label{fig:rise-optics}
RISE optical layout.  The position of the folding mirror is 
also indicated.
}
\end{figure}

\begin{figure}
\begin{center}
\includegraphics[height=8cm]{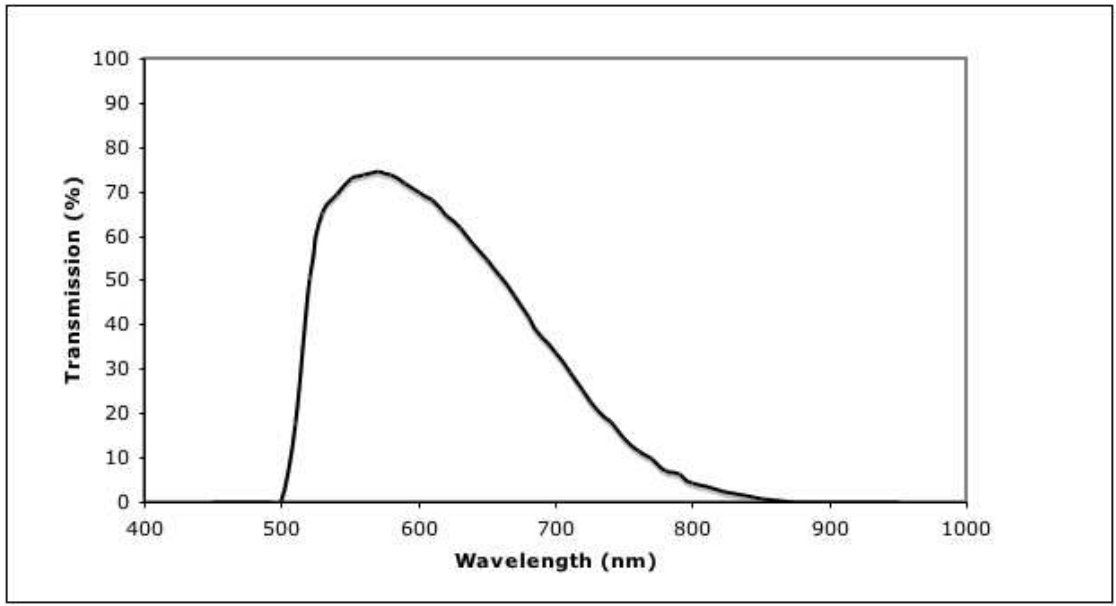}
\end{center}
\caption[]{\label{fig:rise-filter}
Transmission curve for the RISE custom filter constructed from 
2mm KG5 + 3mm OG515.
}
\end{figure}

The overall aim of the optical design (Figure \ref{fig:rise-optics}) of the field condensing chain 
was therefore to achieve a 7.4 arcmin diameter unvignetted field-of-view 
with imaging quality better than the CCD pixel size (0.48 arcsec diameter) 
when projected on to the sky. This is achieved with simple off-the-shelf 
optics which includes a 500 mm focal length, 63 mm diameter, Melles Griot 
``Optimized Achromat'' as a collimator and a Nikkor 135 mm focal length f/2D 
commercial compound lens as the camera. A Newport ``Precision 
Achromatic Doublet'' field lens, just inside the telescope's focal plane, 
places the telescope's exit pupil on to a filter between the collimator 
and camera.  This filter is of a similar design to that built for
the LT RINGO polarimeter\cite{ringo}, in this case 
constructed from 2mm Schott KG5 bonded to 3mm Schott OG515
(Figure \ref{fig:rise-filter}) 

The transmission surfaces of the two achromatic lenses are coated 
with three-layer anti-reflection coatings to minimize losses and ghosts. 
After the field lens the beam is reflected by a 45 degree plane mirror 
in an adjustable holder to permit the field-of-view to be centred precisely 
on the CCD in the camera's focal plane as well as permitting the exit pupil, 
after the collimator, to be centred on the optical axis. It
was found in commissioning (Section \ref{sec:comm}) that the optical aims 
were easily achieved, in fact the useful field-of-view approaches 9 arcmin 
diameter even though vignetted somewhat beyond the design value of 7.4 arcmin.

\subsection{Mechanical Design}

\begin{figure}
\begin{center}
\includegraphics[height=10cm]{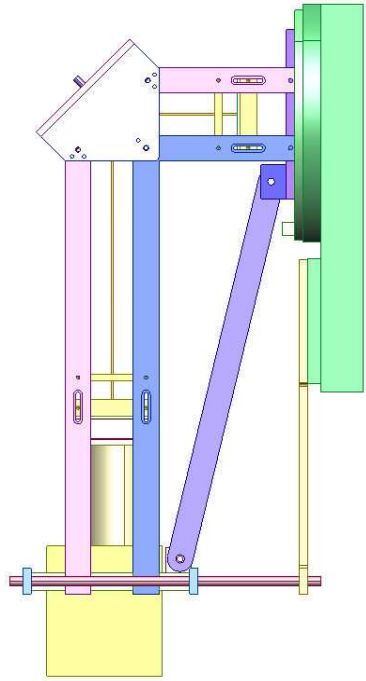}
\end{center}
\caption[]{\label{fig:complete4}
Side view of RISE mechanical design.
}
\end{figure}

\begin{figure}
\begin{center}
\includegraphics[height=10cm]{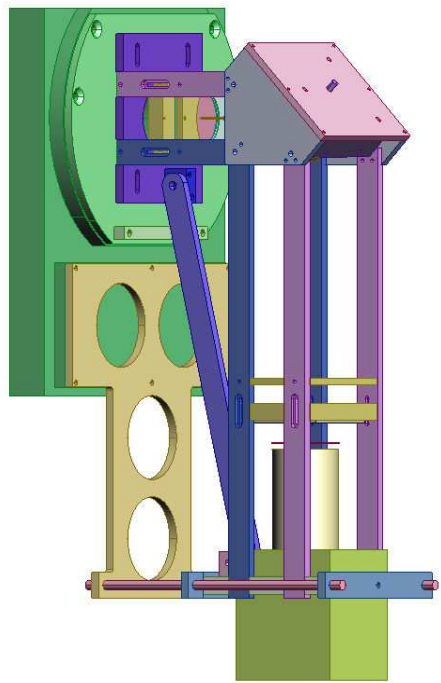}
\end{center}
\caption[]{\label{fig:complete5}
Angle view of RISE mechanical design.
}
\end{figure}

The mechanical design of the instrument (Figures \ref{fig:complete4} 
\& \ref{fig:complete5}) was driven by the
off-the-shelf nature of the optics used. The collimator lens 
has a 500mm focal length; this made the 
instrument too long to be stable without a fold mirror. The fold
mirror brings the optical axis of the instrument alongside the 
Acquisition and Guidance box of the telescope, thereby affording further 
mechanical brace points.  A diagonal bar was fitted spanning the 
length of the instrument to eliminate sag through the 90 degree 
fold. The bottom of the instrument was then clamped to lower part of the 
A\&G box to eliminate `sway'. 
Due to the physical size of this instrument weight was going to be a 
concern, so it was decided to use Aluminum Alloy 5083 H0 for its
light weight yet strong tensile strength. 

There are two lenses in this instrument, a field lens (Newport PAC096)
and a collimator lens (Melles Griot LAO 346). The field lens is 
mounted approximately 10mm inside the focal plane of the telescope 
in an aluminum block, using an O ring mounting to allow 
for the effects of temperature changes at site. The position of this 
aluminum block can be adjusted +/- 15mm.
Then a fold mirror is mounted 150mm from the telescope focal plane, 
this mirror has a purpose built kinematic mount and can be adjusted 
about its centre point easily after the entire instrument has been
assembled, facilitating alignment.
Mounted 350mm further along the optical path is the collimator lens 
in a similar mount to the field lens with the same amount of adjustment
to allow the focus to be set accurately.
Finally the Camera (Andor DW435) with bayonet mounted camera lens 
(Nikkor 135mm) is mounted within a Tufnol plate to electrically 
insulate it from the rest of the instrument.

\subsection{CCD system}

\begin{figure}
\begin{center}
\includegraphics[height=10cm]{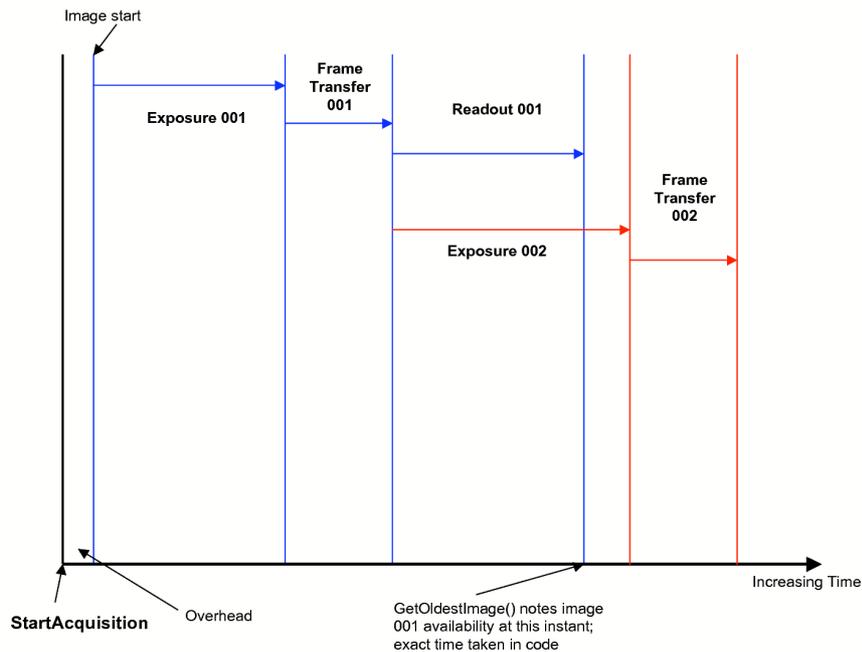}
\end{center}
\caption[]{\label{fig:ian01}
Readout time is $\sim280 {\rm msec}$ and frame transfer time is 
$\sim8 {\rm msec}$ for 2x2 binning at
optimal readout speeds.
}
\end{figure}

The Andor CCD camera was
connected to an Instrument Control Computer (ICC) via a standard PCI connection.
Power for the CCD system is drawn from this connection via a supplemental power
connection to the PCI card from the host system.  
The CCD is constructed with
a light insensitive (LISR) 1024x1024 pixel region adjacent to the LSR pixel
array of the CCD. This acts as a buffer to hold data being moved from the LSR. 
This is the frame-transfer aspect of the CCD, and means that when one image
acquisition has completed on the LSR, it is immediately shunted to the LISR. This
allows another image to start acquiring very quickly, hence the increased time
resolution of the camera. This is in contrast to a standard CCD camera, where
the image must be read out before another is taken. Once an acquisition has
started a series of images is acquired by the camera resulting in an
exposure-frame transfer-readout cycle (Figure \ref{fig:ian01}.)

\begin{figure}
\begin{center}
\includegraphics[height=5cm]{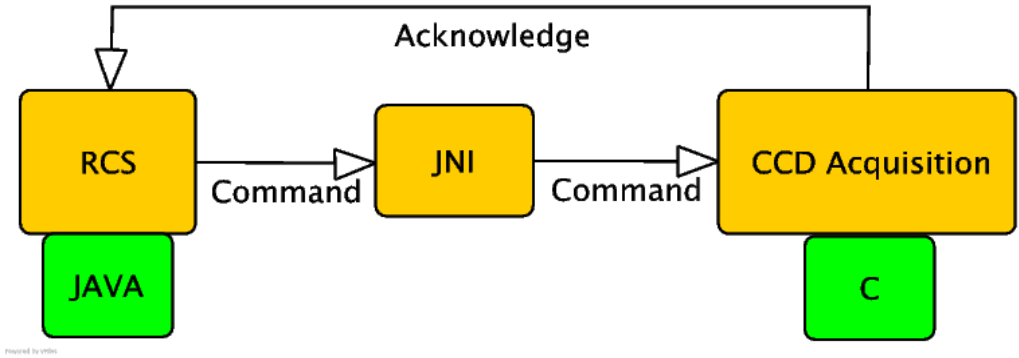}
\end{center}
\caption[]{\label{fig:ian02}
Software command and acknowledgment flow.
}
\end{figure}

The software setup for the RISE camera is broken down into two components: the CCD
control system and the CCD acquisition system. The CCD image 
acquisition system (IAS) is
responsible for instructing the CCD hardware to acquire images, and control
temperature and onboard acquisition characteristics. This is written in C and
makes use of the Andor Linux SDK, allowing camera control software to be written
relatively quickly. The camera control links the C-level acquisition
system to the Java based robotic commands system, which issues the acquisition
commands e.g. calibration set (Figure \ref{fig:ian02}). The Java and C implementations
are linked by the
Java Native Interface (JNI), allowing the Java based Robotic
Control System (RCS) to call the C-based acquisition routines. 

Due to the nature of the image acquisition software the data rate is very high
for short exposures. Data can be acquired in 1x1 or 2x2 binning modes, with
minimum exposure cycle times of $1.8$ sec and $1.0$ sec respectively; thus the
maximum data generation capacity can be $\sim4.0\,$Gbytes per hour for a 1x1 data run or
$\sim3.6\,$Gbytes per hour in 2x2 mode. Another side effect of the high data is
the need to run the acquisition process in isolation from the the ICS. The
usual LT method of acquisition is to take an image and return an
acknowledgment to the RCS between each image. 
However, to do so in this case would create timing delays and
introduce an unusual level of packet traffic between the IAS and the RCS. The
solution to this is to yield control of the instrument to the IAS and send an
acknowledge at the end of a long run. 

Bias, flat and science frames are taken through two modes of acquisition: {\it
twilight calibrate} and {\it multrun}. Since there is no shutter on the CCD, it
can be difficult to acquire a bias image. This must be achieved manually by
taking zero second exposures with the light path to the CCD obstructed. Dark
frames are taken in a similar fashion with the `multrun command'. Flats
are taken via the `twilight calibrate' command which adapts the
exposure time to achieve a similar number of counts on each flat image
based on the median count in the previous image. Finally,
the `multrun' command takes a series of images in frame transfer mode. The
number of images required is set along with the exposure time. The acquired data is
stored locally on the ICC before being transferred back to LJMU next day, where
it is made available via staging on a webserver.  

\section{Commissioning}

\label{sec:comm}

RISE was fitted to the Liverpool Telescope in February 2008.
Setup was relatively straight forward. The instrument was fitted 
to the telescope without the CCD camera. A test card was clamped to the 
field lens holder and the holder was adjusted to the focal point of 
the telescope. A 1 metre focal length telescope (set to infinity) 
was then positioned where the camera would eventually sit and the 
collimator lens was adjusted until the test card could be viewed in 
focus. The test card was removed and the field lens adjusted away 
from the focal plane to keep any dust particles out of focus. The 
mirror was then roughly adjusted by eye to bring all the optical 
components concentric within the instrument.  The camera was then fitted 
and images taken of the inside of the enclosure.  A combination
of small adjustments of the telescope fold mirror in the A\&G box and the
fold mirror within the instrument itself were then made
to centre the image.

\begin{figure}
\begin{center}
\includegraphics[height=9.8cm]{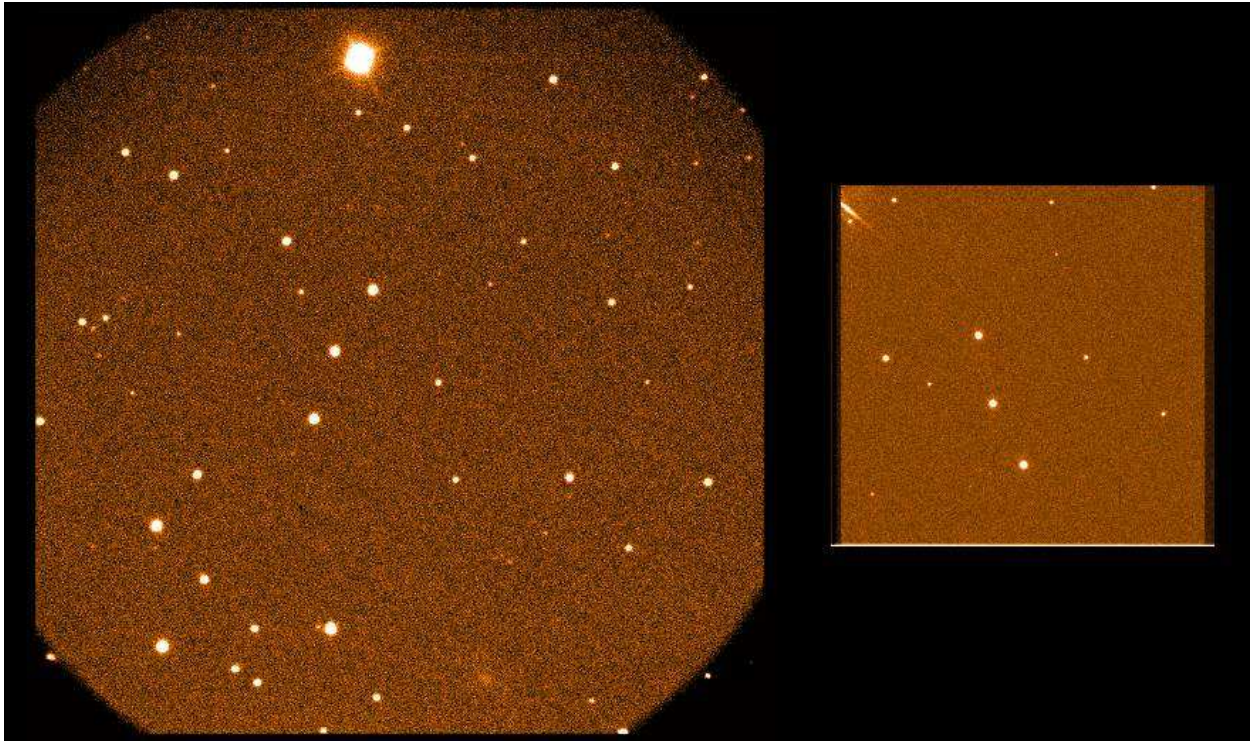}
\end{center}
\caption[]{\label{fig:rise-ratcam}
A comparison of raw images from the RISE (left hand side)
and RATCam (right hand side) cameras of the Liverpool
Telescope to the same scale.  
}
\end{figure}

On sky commissioning took place over the period 20-23rd February 2008.
The field of view (with some vignetting in the corners) was found
to be 9.4 x 9.4 arcminutes, corresponding to a pixel scale of 
0.55 arcsec/pixel (Figure \ref{fig:rise-ratcam}).  
No significant variation in point spread function across 
the field could be detected.  The detector system read-noise was measured
at 10 electrons rms, and the gain as 2.4 electrons/count.  Linearity
testing showed good performance up to the ADC conversion limit
to 65,000 counts with 1x1 binning and around 40,000 counts
with 2x2 binning.  With the short exposure times used for the
instrument no significant dark current could be detected.

\section{data reduction}

\begin{figure}
\begin{center}
\includegraphics[height=21cm]{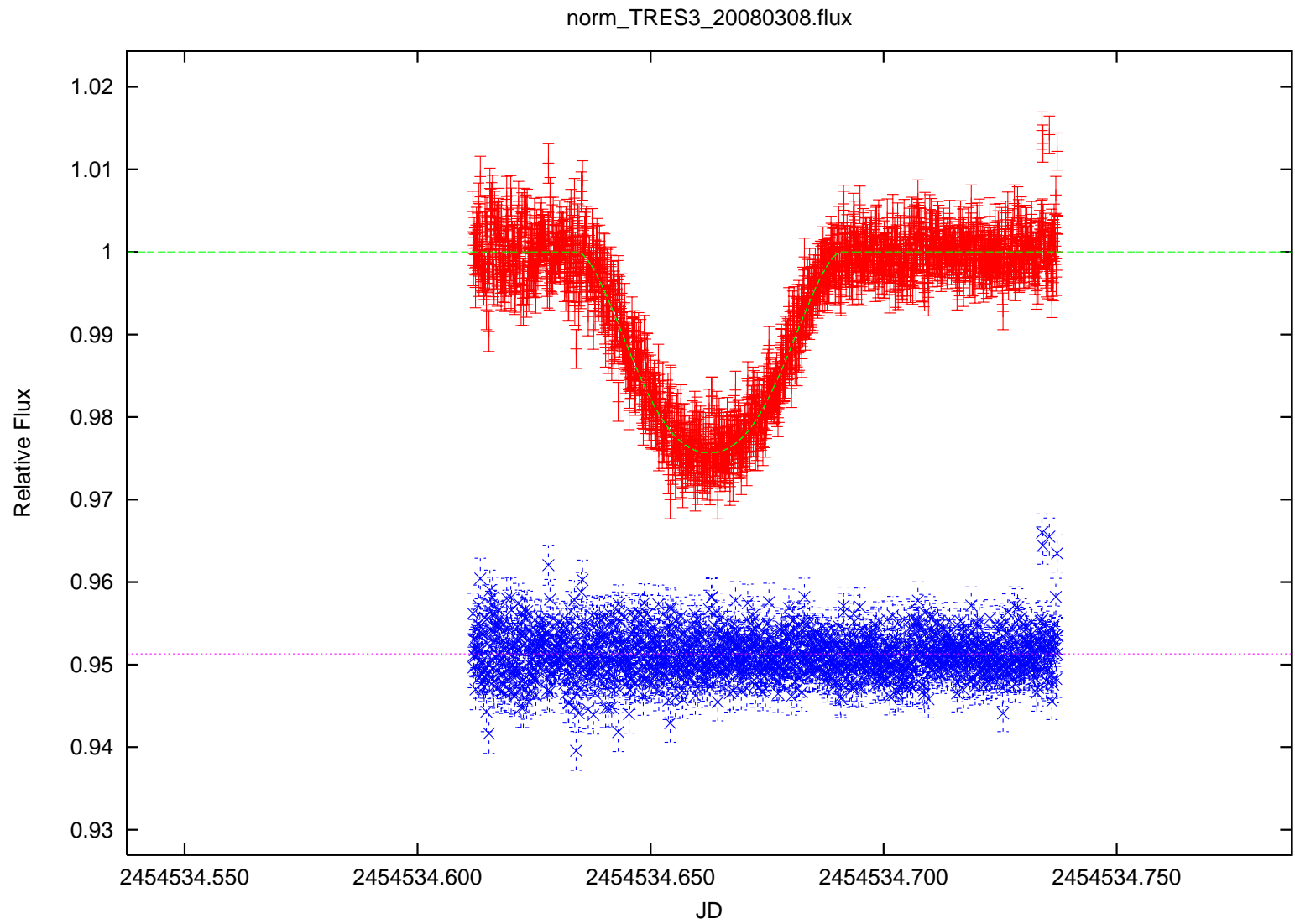}
\end{center}
\caption[]{\label{fig:TRES3}
Light curve and model residuals for the transiting exoplanet TRES3.
}
\end{figure}

The reduction of data to milli-mag accuracies is a complex
task, and requires a good understanding of the systematic errors.
At present our data reduction procedures are still at an experimental 
stage, although we can already demonstrate high quality photometry
and timing results (Figure \ref{fig:TRES3}).  

\begin{figure}
\begin{center}
\includegraphics[height=20cm]{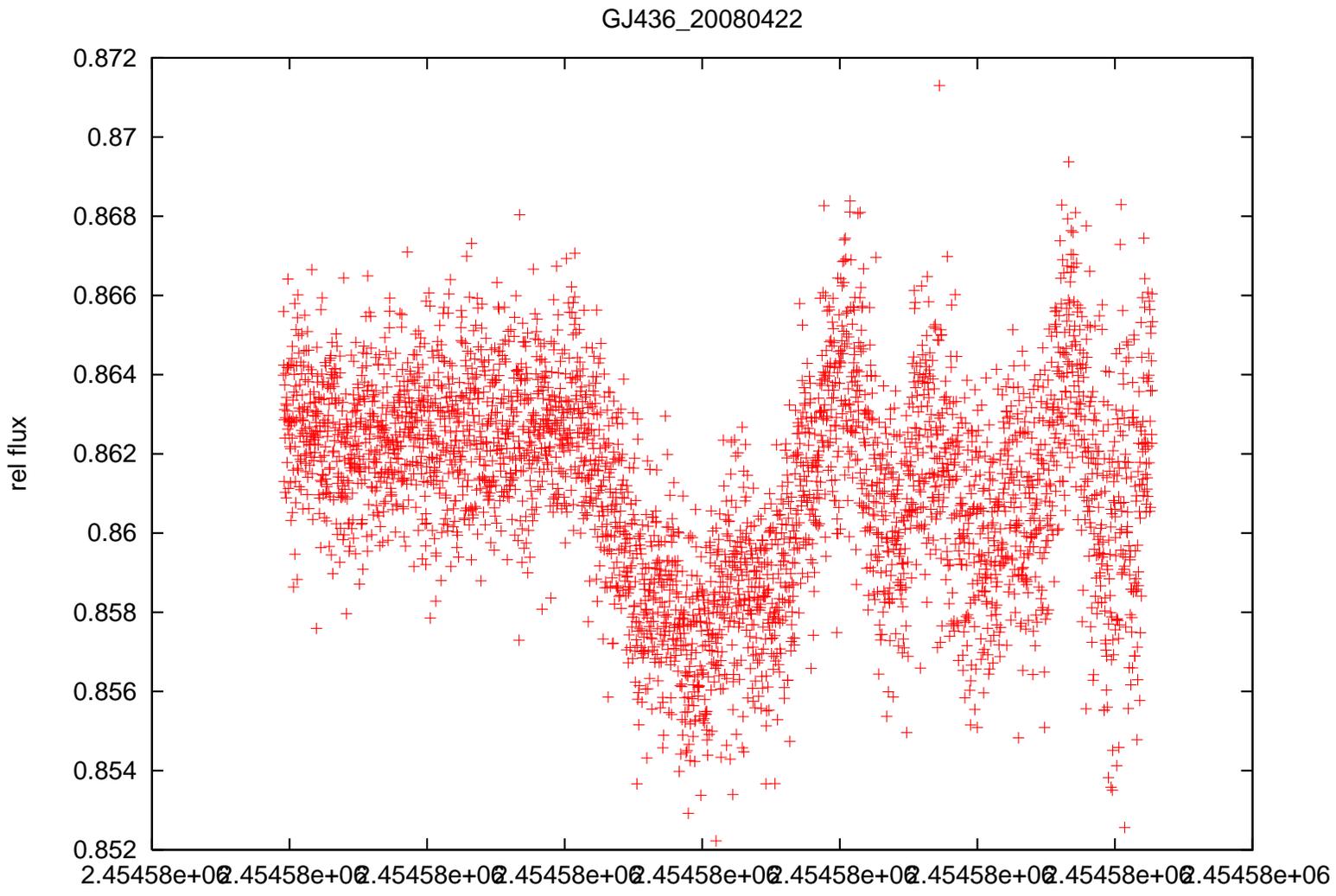}
\end{center}
\caption[]{\label{fig:GL436}
RISE lightcurve of GJ436.  Note how the loss of the autoguider
half way through the sequence of exposures and the resulting
drift of the target across the CCD causes problems with the
photometry.  Before loss of the autoguider however, the eclipse
(with a depth of 0.006 magnitudes) is easily visible.
}
\end{figure}

\begin{figure}
\begin{center}
\includegraphics[height=10cm]{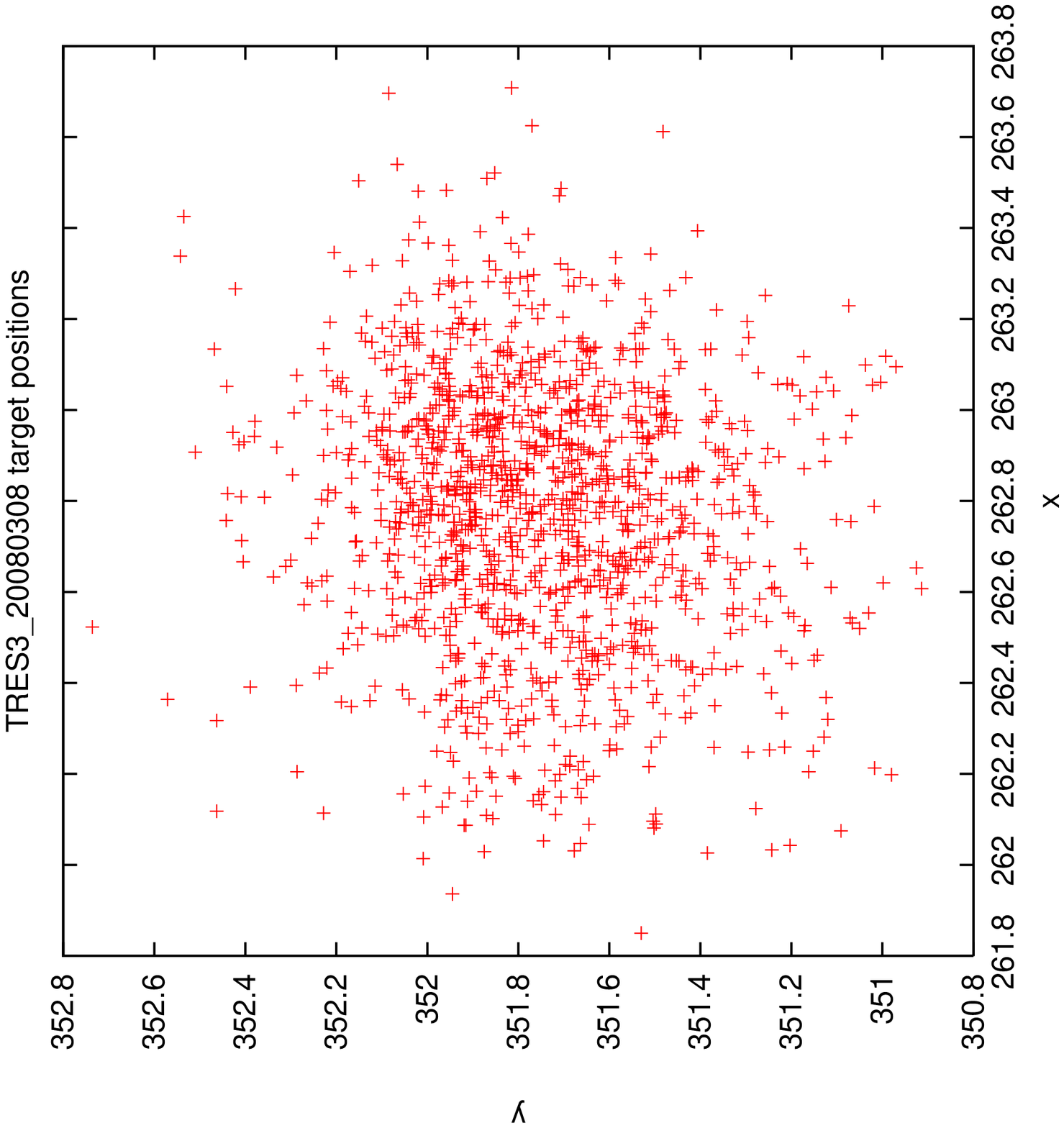}

\includegraphics[height=10cm]{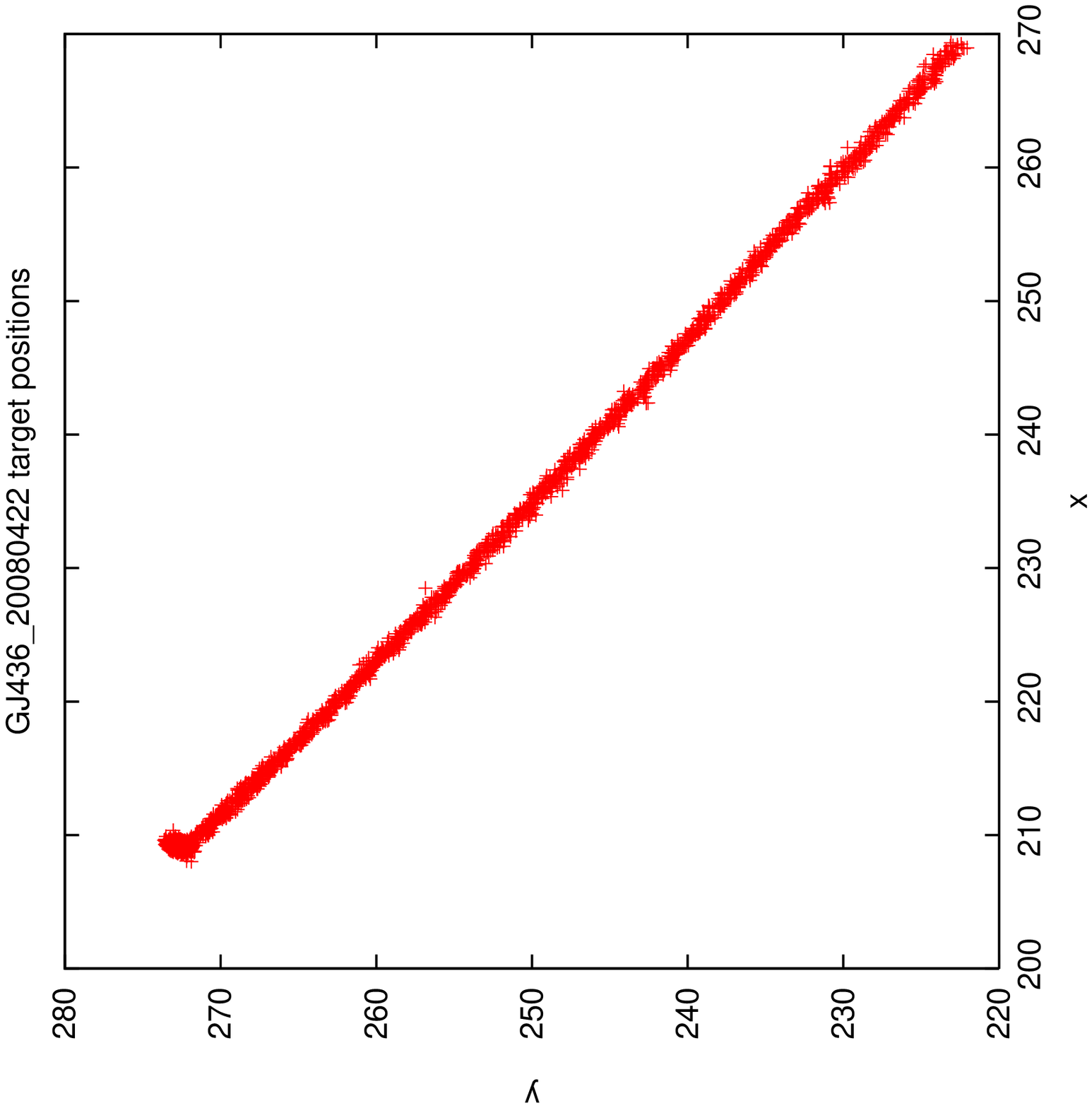}
\end{center}
\caption[]{\label{fig:autoguider}
Targets star centroids for the observations of TRES3 (autoguided) and
GL436 (not autoguided).
}
\end{figure}

The initial stages the raw images are first 
put through a python/pyraf script that updates 
the headers that are needed at the reduction or lightcurve fitting stage, 
including the heliocentric Julian date and airmass (calculated from DATE-OBS, 
RA and DEC), plus other instrument characteristics such as the gain, read-noise and 
filter. Due to the instrument not  having a 
shutter we try both bias images and bias 
strips to debias the images. Flat fielding is also tricky. We currently do not 
understand how the flat field structure changes during the course of a night (due to scattered light in the telescope and optics) and 
therefore we prefer not to flat field 
in cases where autoguiding is stable. This is because the ratio of 
the stars fluxes on the chip will stay constant and flat fielding effects
will be removed to first order during sky subtraction anyway. If however
there is some drift (e.g. due to lack of an appropriate guide star) 
flat fielding must be used to remove some (but not all) of the systematic effects. 
An example of this is shown in Figure \ref{fig:GL436} where autoguiding was
lost half way through the run (Figure \ref{fig:autoguider}) - the 
detrimental effect on
the light-curve is obvious.  

Aperture photometry is performed on the images using IRAF/daophot via a
python/pyraf script that uses a list of apertures to calculate the relative 
flux of the target star (given a list of reference stars coordinates)
and the respective errors. Typically an aperture of 10 pixels is 
used (in 2x2 binning mode).   Following this the lightcurves are then 
normalized using an airmass (or time) function
and a model is fitted to the lightcurves according to an analytic 
formula\cite{mandel02} using a Monte Carlo Markov Chain code with
suitable priors on known stellar and orbital parameters.   This allows
us to calculate the radius, inclination and central transit times.  It is
the variation of these central transit times from the predicted ephemeris that
will provide evidence of a third body.

\section{CONCLUDING REMARKS}

RISE was a fast-track instrument for the
Liverpool Telescope, which was designed, built and
commissioned within 1 year of the initial science
requirement being identified. 
Although work
is still underway to better understand the
effect of scattered light on the observations,
it is already clear that the timing
precision obtainable 
is sufficient to carry out our programme
of transit followup, which has therefore begun routine 
data taking.  In addition we are pleased to note that 
it is also attracting interest from
other science users of the telescope attracted by its low
overheads for timing experiments and its relatively
wide field.

%%%%%%%%%%%%%%%%%%%%%%%%%%%%%%%%%%%%%%%%%%%%%%%%%%%%%%%%%%%%%
\acknowledgments     %>>>> equivalent to \section*{ACKNOWLEDGMENTS}       
 
We thank the staff of the QUB Physics workshop for their
excellent work in constructing RISE.  The Liverpool Telescope
is operated on the island of La Palma by Liverpool John Moores
University in the Spanish Observatorio del Roque de los Muchachos
of the Instituto de Astrofisica de Canaries with financial
support from the UK Science and Technologies Facilities
Council. 

%%%%%%%%%%%%%%%%%%%%%%%%%%%%%%%%%%%%%%%%%%%%%%%%%%%%%%%%%%%%%
%%%%% References %%%%%

\bibliography{rise}   %>>>> bibliography data in report.bib

\begin{thebibliography}{10}

\bibitem{steele}
{Steele}, I.~A., {Smith}, R.~J., {Rees}, P.~C., {Baker}, I.~P., {Bates}, S.~D.,
  {Bode}, M.~F., {Bowman}, M.~K., {Carter}, D., {Etherton}, J., {Ford}, M.~J.,
  {Fraser}, S.~N., {Gomboc}, A., {Lett}, R.~D.~J., {Mansfield}, A.~G.,
  {Marchant}, J.~M., {Medrano-Cerda}, G.~A., {Mottram}, C.~J., {Raback}, D.,
  {Scott}, A.~B., {Tomlinson}, M.~D., and {Zamanov}, R., ``{The Liverpool
  Telescope: performance and first results},'' in [{\em Proceedings of the
  SPIE, Volume 5489, pp. 679-692 (2004).}{\nolinebreak\hspace{0.1em}]},
  {Oschmann}, J.~M., ed.,  679--692 (Oct. 2004).

\bibitem{mottram}
{Mottram}, C.~J., {Steele}, I.~A., and {Morales}, L., ``{Design of low cost and
  reliable instrumentation for robotic telescopes},'' in [{\em Ground-based
  Instrumentation for Astronomy. Edited by Alan F. M. Moorwood and Iye
  Masanori. Proceedings of the SPIE, Volume 5492, pp. 677-688
  (2004).}{\nolinebreak\hspace{0.1em}]},  {Moorwood}, A.~F.~M. and {Iye}, M.,
  eds.,  677--688 (Sept. 2004).

\bibitem{fraser}
{Fraser}, S. and {Steele}, I.~A., ``{Robotic telescope scheduling: the
  Liverpool Telescope experience},'' in [{\em Optimizing Scientific Return for
  Astronomy through Information Technologies. Edited by Quinn, Peter J.;
  Bridger, Alan. Proceedings of the SPIE, Volume 5493, pp. 331-340
  (2004).}{\nolinebreak\hspace{0.1em}]},  {Quinn}, P.~J. and {Bridger}, A.,
  eds., {\em Presented at the Society of Photo-Optical Instrumentation
  Engineers (SPIE) Conference} {\bf 5493},  331--340 (Sept. 2004).

\bibitem{robonet}
{Burgdorf}, M.~J., {Bramich}, D.~M., {Dominik}, M., {Bode}, M.~F., {Horne},
  K.~D., {Steele}, I.~A., {Rattenbury}, N., and {Tsapras}, Y., ``{Exoplanet
  detection via microlensing with RoboNet-1.0},'' {\em PLANSS}~{\bf 55},
  582--588 (Apr. 2007).

\bibitem{planet}
{Gaudi}, B.~S., {Bennett}, D.~P., {Udalski}, A., {Gould}, A., {Christie},
  G.~W., {Maoz}, D., {Dong}, S., {McCormick}, J., {Szyma{\'n}ski}, M.~K.,
  {Tristram}, P.~J., {Nikolaev}, S., {Paczy{\'n}ski}, B., {Kubiak}, M.,
  {Pietrzy{\'n}ski}, G., {Soszy{\'n}ski}, I., {Szewczyk}, O., {Ulaczyk}, K.,
  {Wyrzykowski}, {\L}., {DePoy}, D.~L., {Han}, C., {Kaspi}, S., {Lee}, C.-U.,
  {Mallia}, F., {Natusch}, T., {Pogge}, R.~W., {Park}, B.-G., {Abe}, F.,
  {Bond}, I.~A., {Botzler}, C.~S., {Fukui}, A., {Hearnshaw}, J.~B., {Itow}, Y.,
  {Kamiya}, K., {Korpela}, A.~V., {Kilmartin}, P.~M., {Lin}, W., {Masuda}, K.,
  {Matsubara}, Y., {Motomura}, M., {Muraki}, Y., {Nakamura}, S., {Okumura}, T.,
  {Ohnishi}, K., {Rattenbury}, N.~J., {Sako}, T., {Saito}, T., {Sato}, S.,
  {Skuljan}, L., {Sullivan}, D.~J., {Sumi}, T., {Sweatman}, W.~L., {Yock},
  P.~C.~M., {Albrow}, M.~D., {Allan}, A., {Beaulieu}, J.-P., {Burgdorf}, M.~J.,
  {Cook}, K.~H., {Coutures}, C., {Dominik}, M., {Dieters}, S., {Fouqu{\'e}},
  P., {Greenhill}, J., {Horne}, K., {Steele}, I., {Tsapras}, Y., {Chaboyer},
  B., {Crocker}, A., {Frank}, S., and {Macintosh}, B., ``{Discovery of a
  Jupiter/Saturn Analog with Gravitational Microlensing},'' {\em Science}~{\bf
  319},  927-- (2008).

\bibitem{holman05}
{Holman}, M.~J. and {Murray}, N.~W., ``{The Use of Transit Timing to Detect
  Terrestrial-Mass Extrasolar Planets},'' {\em Science}~{\bf 307},  1288--1291
  (Feb. 2005).

\bibitem{algol05}
{Agol}, E., {Steffen}, J., {Sari}, R., and {Clarkson}, W., ``{On detecting
  terrestrial planets with timing of giant planet transits},'' {\em MNRAS}~{\bf
  359},  567--579 (May 2005).

\bibitem{wasp1}
{Cameron}, A.~C., {Bouchy}, F., {H{\'e}brard}, G., {Maxted}, P., {Pollacco},
  D., {Pont}, F., {Skillen}, I., {Smalley}, B., {Street}, R.~A., {West}, R.~G.,
  {Wilson}, D.~M., {Aigrain}, S., {Christian}, D.~J., {Clarkson}, W.~I.,
  {Enoch}, B., {Evans}, A., {Fitzsimmons}, A., {Fleenor}, M., {Gillon}, M.,
  {Haswell}, C.~A., {Hebb}, L., {Hellier}, C., {Hodgkin}, S.~T., {Horne}, K.,
  {Irwin}, J., {Kane}, S.~R., {Keenan}, F.~P., {Loeillet}, B., {Lister}, T.~A.,
  {Mayor}, M., {Moutou}, C., {Norton}, A.~J., {Osborne}, J., {Parley}, N.,
  {Queloz}, D., {Ryans}, R., {Triaud}, A.~H.~M.~J., {Udry}, S., and {Wheatley},
  P.~J., ``{WASP-1b and WASP-2b: two new transiting exoplanets detected with
  SuperWASP and SOPHIE},'' {\em MNRAS}~{\bf 375},  951--957 (Mar. 2007).

\bibitem{wasp3}
{Pollacco}, D., {Skillen}, I., {Collier Cameron}, A., {Loeillet}, B.,
  {Stempels}, H.~C., {Bouchy}, F., {Gibson}, N.~P., {Hebb}, L., {H{\'e}brard},
  G., {Joshi}, Y.~C., {McDonald}, I., {Smalley}, B., {Smith}, A.~M.~S.,
  {Street}, R.~A., {Udry}, S., {West}, R.~G., {Wilson}, D.~M., {Wheatley},
  P.~J., {Aigrain}, S., {Alsubai}, K., {Benn}, C.~R., {Bruce}, V.~A.,
  {Christian}, D.~J., {Clarkson}, W.~I., {Enoch}, B., {Evans}, A.,
  {Fitzsimmons}, A., {Haswell}, C.~A., {Hellier}, C., {Hickey}, S., {Hodgkin},
  S.~T., {Horne}, K., {Hrudkov{\'a}}, M., {Irwin}, J., {Kane}, S.~R., {Keenan},
  F.~P., {Lister}, T.~A., {Maxted}, P., {Mayor}, M., {Moutou}, C., {Norton},
  A.~J., {Osborne}, J.~P., {Parley}, N., {Pont}, F., {Queloz}, D., {Ryans}, R.,
  and {Simpson}, E., ``{WASP-3b: a strongly irradiated transiting gas-giant
  planet},'' {\em MNRAS}~{\bf 385},  1576--1584 (Apr. 2008).

\bibitem{wasp4}
{Wilson}, D.~M., {Gillon}, M., {Hellier}, C., {Maxted}, P.~F.~L., {Pepe}, F.,
  {Queloz}, D., {Anderson}, D.~R., {Collier Cameron}, A., {Smalley}, B.,
  {Lister}, T.~A., {Bentley}, S.~J., {Blecha}, A., {Christian}, D.~J., {Enoch},
  B., {Haswell}, C.~A., {Hebb}, L., {Horne}, K., {Irwin}, J., {Joshi}, Y.~C.,
  {Kane}, S.~R., {Marmier}, M., {Mayor}, M., {Parley}, N., {Pollacco}, D.,
  {Pont}, F., {Ryans}, R., {Segransan}, D., {Skillen}, I., {Street}, R.~A.,
  {Udry}, S., {West}, R.~G., and {Wheatley}, P.~J., ``{WASP-4b: A 12th
  Magnitude Transiting Hot Jupiter in the Southern Hemisphere},'' {\em
  ApJ}~{\bf 675},  L113--L116 (Mar. 2008).

\bibitem{wasp5}
{Anderson}, D.~R., {Gillon}, M., {Hellier}, C., {Maxted}, P.~F.~L., {Pepe}, F.,
  {Queloz}, D., {Wilson}, D.~M., {Collier Cameron}, A., {Smalley}, B.,
  {Lister}, T.~A., {Bentley}, S.~J., {Blecha}, A., {Christian}, D.~J., {Enoch},
  B., {Hebb}, L., {Horne}, K., {Irwin}, J., {Joshi}, Y.~C., {Kane}, S.~R.,
  {Marmier}, M., {Mayor}, M., {Parley}, N.~R., {Pollacco}, D.~L., {Pont}, F.,
  {Ryans}, R., {S{\'e}gransan}, D., {Skillen}, I., {Street}, R.~A., {Udry}, S.,
  {West}, R.~G., and {Wheatley}, P.~J., ``{WASP-5b: a dense, very hot Jupiter
  transiting a 12th-mag Southern-hemisphere star},'' {\em MNRAS} ,  L49+ (Apr.
  2008).

\bibitem{ringo}
{Steele}, I.~A., {Bates}, S.~D., {Carter}, D., {Clarke}, D., {Gomboc}, A.,
  {Guidorzi}, C., {Melandri}, A., {Monfardini}, A., {Mottram}, C.~J.,
  {Mundell}, C.~G., {Scott}, A.~B., {Smith}, R.~J., and {Swindlehurst}, J.,
  ``{RINGO: a novel ring polarimeter for rapid GRB followup},'' in [{\em
  Ground-based and Airborne Instrumentation for Astronomy. Edited by McLean,
  Ian S.; Iye, Masanori. Proceedings of the SPIE, Volume 6269, pp. 62695M
  (2006).}{\nolinebreak\hspace{0.1em}]},  {\em Presented at the Society of
  Photo-Optical Instrumentation Engineers (SPIE) Conference} {\bf 6269} (July
  2006).

\bibitem{mandel02}
{Mandel}, K. and {Agol}, E., ``{Analytic Light Curves for Planetary Transit
  Searches},'' {\em ApJ}~{\bf 580},  L171--L175 (Dec. 2002).

\end{thebibliography}
\bibliographystyle{spiebib}   %>>>> makes bibtex use spiebib.bst

\end{document}